# A Novel FPGA-based CNN Hardware Accelerator: Optimization for Convolutional Layers using Karatsuba Ofman Multiplier

Amit Sarkar
*Graduate Student IEEE Member*
sarkaramit424@gmail.com

*Abstract*—A new architecture of CNN hardware accelerator is presented. Convolutional Neural Networks (CNNs) are a subclass of neural networks that have demonstrated outstanding performance in a variety of computer vision applications, including object detection, image classification, and many more. Convolution, a mathematical operation that consists of multiplying, shifting and adding a set of input values by a set of learnable parameters known as filters or kernels, which is the fundamental component of a CNN. The Karatsuba Ofman multiplier is known for its ability to perform high-speed multiplication with less hardware resources compared to traditional multipliers. This article examines the usage of the Karatsuba Ofman Multiplier method on FPGA in the prominent CNN designs AlexNet, VGG16, and VGG19.

## I. INTRODUCTION

Convolution is a mathematical operation that involves multiplying, shifting and addition of two functions, typically an image or signal and a kernel. In the context of CNNs, convolution is used to extract features from an image. When an input image is convolved with a kernel, it generates a set of output feature maps. The kernel consists of learnable parameters. A lot of signal processing applications use FPGA-based CNN hardware accelerators [1-4] because of their programmability, flexibility, and high performance. The memory bottleneck, however, is one of the main issues that FPGA-based CNN hardware accelerators must deal with. In CNNs, the network's weights and biases are stored in memory and accessed during its forward and backward propagation phases. The amount of memory needed to hold a network's weights and biases grows as the network gets bigger, and for large networks, the memory requirements might be substantial. To implement CNNs on FPGA custom hardware architectures for convolution, pooling, and fully connected operations must be developed. Since convolution is the most significant operation in CNNs, it is essential to optimize resource utilization in this operation. Convolutional layers are usually implemented on FPGAs using the systolic array architecture, which stores the input and weights in memory and performs multiplication and accumulation in a pipelined manner using registers. Specialized hardware architectures like average-pooling or max-pooling can be used to implement pooling layers on FPGAs. Fully connected layers that require matrix-vector multiplication can also be achieved on FPGAs by utilizing hardware with matrix multiplication-optimized topologies. FPGAs can be combined with traditional processors to create hybrid computing systems, where CPUs handle less computationally intensive tasks, and FPGAs handle the computationally intensive ones. Several CNNs exist, including Alexnet[5], VGG16 [6], and VGG19 [6]. VGG16 and VGG19 both feature 12 and 14 convolutional layers, compared to Alexnet's 5 layers. VGG16 and VGG19 have images that are of size 224x224x3, while Alexnet has an image that is of size 227x227x3. VGG16 and VGG19 each have 3968 3x3 kernel matrices and 4992 3x3 kernel matrices, respectively. The Alexnet network, on the other hand, includes 1024 3x3 kernel matrices, 256 5x5 kernel matrices, and 96 11x11 kernel matrices. The paper is organized as follows: Section 2 and 3 presents the proposed architecture of CNN Hardware Accelerator. Section 4 discusses the Karatsuba Ofman Multiplier. The result and analysis are provided in Section 5. And finally, the conclusion in Section 6.

## II. ARCHITECTURE OF CNN HARDWARE ACCELERATOR

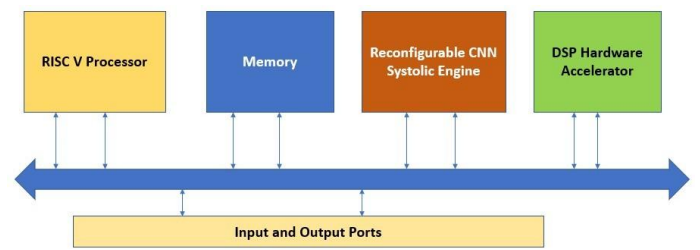

Fig. 1. Architecture of CNN Hardware Accelerator

Machine intelligence is necessary for a range of signal processing applications, including image recognition, speech recognition, robot navigation, cognitive radio, and target detection. However, integrating high-performance signal processing engines and AI engines that use deep neural networks on the same chip can be challenging as it requires high performance, low cost, low power consumption, and minimal area usage. Programmable devices can be adaptive to specific user needs, but hard-wired chips are inflexible. Whereas Reconfigurable networks allows efficient use of chip resources to build networks with various topologies. A CNN implemented on an FPGA is composed of numerous processing elements systolic cells, each responsible for a distinct task in handling the input data. These systolic cells are interconnected in a pipeline, enabling the input data to be processed efficiently. The systolic cell is equipped with a register that stores the corresponding coefficient (h) retrieved from memory. Each systolic cell is composed of a left-hand input (Yn-1), a vertical input (X(n)), and a right-hand output (Yn). Additionally, this block is fitted with an adder and a multiplier. With every clock pulse, the systolic cell executes and the output is given by:

$Y_n = Y_{n-1} + h.X(n)$

In order to comprehend the use of systolic cells in convolutional neural network, let's examine the configuration of a one-dimensional convolution or the implementation of a Finite Impulse Response (FIR) filter through a systolic array. The

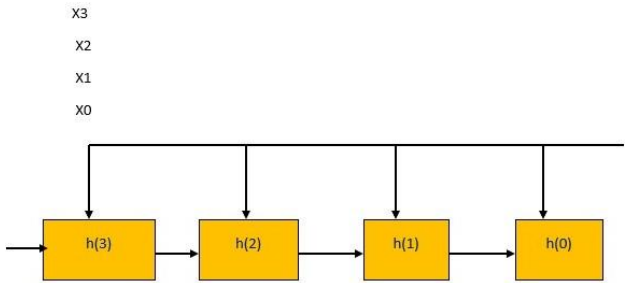

Fig. 2. Systolic Architecture of 1D FIR filter

filter's systolic architecture operates by passing each input sample through multiple processing cells. Each cell conducts a MAC operation on the input signal by multiplying it with filter coefficients stored in the cell and adding it to the output of the previous systolic cell. This process is repeated for each new set of input samples that arrive, resulting in a steady stream of output. The final output is given by

$y[n] = \sum [h(k)x[n-k]]$

Which is equivalent to the output of 1D FIR filter. Additionally primary operation of a neural network is the summation of WiXi that has some bias and processed through an activation function. Where weight and input vector Wi*Xi multiplication is crucial. Systolic cell architecture could easily achieve this by, for example, storing the weight in place of h(n). In the case of the 2D convolution utilised by CNN, multiplication refers to matrix multiplication followed by shifting and adding. Convolution is defined as shifting, multiplying, and adding in the 1D context. For example convolution of 3x3x3 kernel matrix with a 5x5x3 picture matrix produces a 3x3x3 matrix. And based on the systolic implementation of CNN, it can be inferred that multiplication is a crucial algorithm for implementing CNNs on an FPGA. While ASICs offer high performance at low silicon costs, they may not be cost-effective for applications requiring multiple real-time functions. Field Programmable Gate Arrays (FPGAs) are ideal for various signal processing applications as they are programmable hardware that can be dynamically reconfigured at runtime. However, a high number of configurable logic blocks (CLBs) needs to be interconnected, leading to a complex routing structure and interconnection delays. Thus a Reconfigurable Systolic engine is proposed for the CNN hardware acclerator

The proposed design involves a RISC V processor controlling the Reconfigurable Systolic Engine, DSP Hardware Accelerator, and Memory. The Reconfigurable CNN Systolic Engine consists of systolic cells that can be configured to realize different algorithms within the same architecture. The systolic architecture[7-10] is used in various areas such as neural networks, image processing, signal processing, 2D convolution correlation, Discrete Fourier transform, and matrix arithmetic. Previous work has demonstrated the advantages of systolic architecture in speeding up computations, providing modular approaches, minimizing interconnection costs, and overcoming the mismatch between computation and architecture. A RISC V processor can configure the connection between systolic cells to realize various modules for CNN, including convolution, pooling, and fully connected layers.

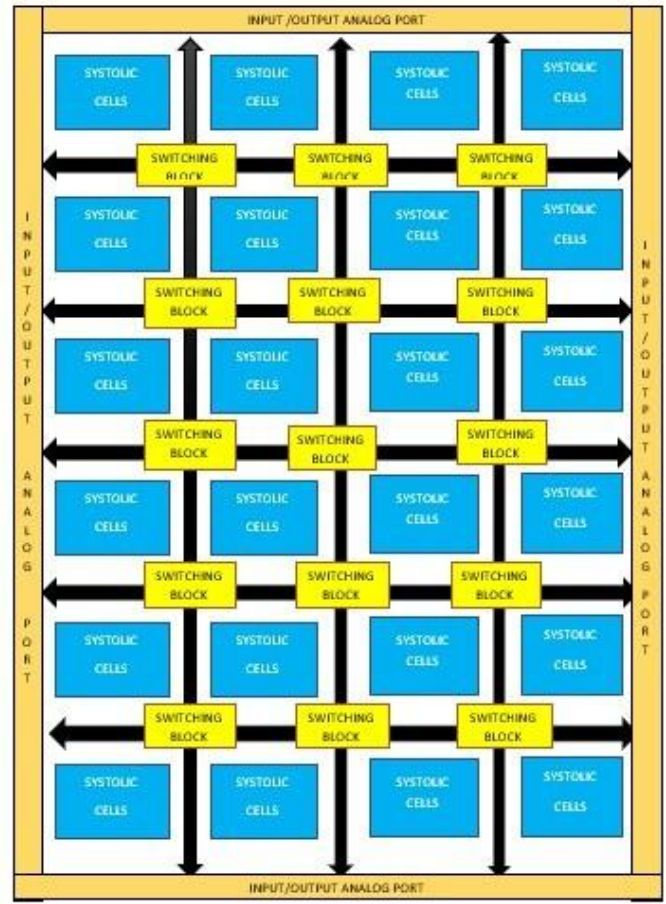

Fig. 3. Architecture of Re-configurable systolic engine

## III. ARCHITECTURE OF RE-CONFIGURABLE SYSTOLIC ENGINE

The Architecture of Reconfigurable systolic engine is shown in figure 3. The working of Reconfigurable Systolic Engine is as follows:

1. The instructions to configure systolic cells or the bit file is stored in a memory.

2. Custom Hardware: Depending upon the type of instruction like FFT, FIR the connection between the systolic cell is configured

3. On FPGA: The bit file for the required architecture is downloaded to FPGA from the memory.

Next, the design of the control unit for Reconfigurable systolic engine is discussed. In this design, the processor will be initially programmed to configure the FPGA-like execution without needing a separate CPU for configuration. The instructions will be stored in the instruction/program memory and used to configure the hardware. The Reconfigurable systolic engine includes a significant number of systolic cell, switches and input/output ports. Depending on the type of CNN module (Ex: Convolution, pooling, fully connected) being used, the hardware will be configured accordingly.

## IV. KARATSUBA OFMAN MULTIPLIER:

A quick algorithm for multiplying two large integers is the Karatsuba Ofman Multiplier [11]. Anatolii Alexeevitch Karatsuba and Yuri Borisovich Ofman developed this in the 1960s. The Karatsuba ofman multiplier uses a divide and conquer algorithm, in which a bigger problem is broken into smaller problems. After that the problems are being computed. Finally, all the results are combined to get the ultimate result. For example, the multiplication between A and B. Where A is the multiplier and B is the multiplicand. It can be represented as

$A*B = (Al*Bl)* 2^n + (Ar*Bl)*2^{n/2} + (Al*Br)*2^{n/2} + Ar*Br$

Al and Bl are left half of A and B. whereas Ar and Br are the right half of A and B. 'n' is the number of bits. This segmentation of the multiplier and multiplicand in both halves continue until each segment become 2-bits. After that the computation is carried out for each segment and they are added together to get the final result. The procedure is very effective when multiplying large integers since it minimises the number of multiplications needed to compute the product. A RTL schematic and simulation result of 32 bit Pipelined High speed Karatsuba Ofman Multiplier is shown in Figure 4 and Figure 5.

## V. RESULT AND ANALYSIS

The analysis shows that the Karatsuba-Ofman multiplier has lower delay compared to other multipliers. In Convolutional Neural Networks, the multiplication of two matrices of the same size is the key operation in the convolutional layer.

This operation requires $n^3$

multipliers for two matrices of size n x n. The number of convolutional layers and kernel matrices differ in Alexnet,

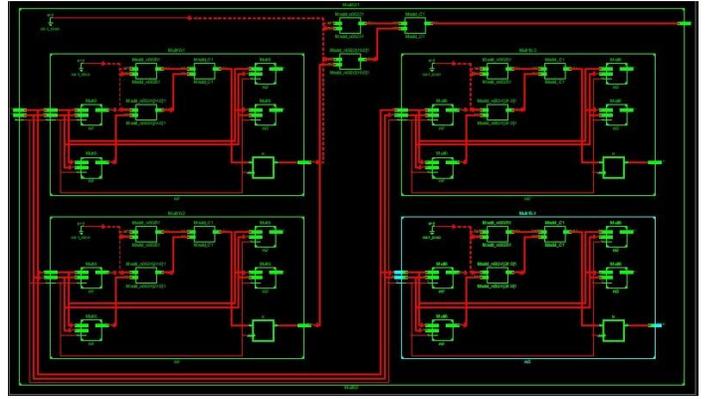

Fig. 4. RTL schematic of 32-bit multiplier

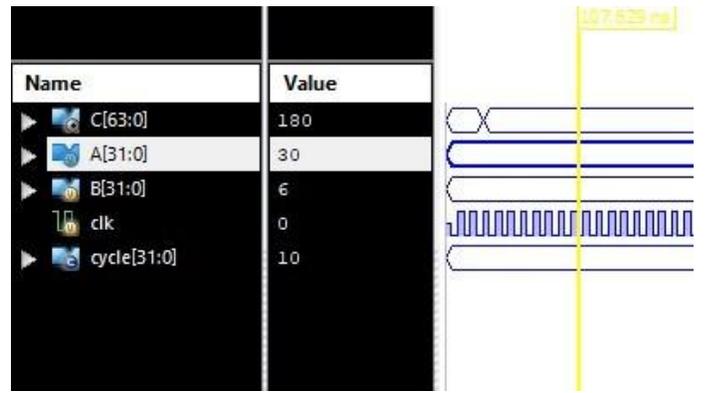

Fig. 5. Simulation Result of 32-bit KOM Multiplier

Table 1: Multiplication of 3x3 with another 3x3 matrix

| Sl no | Logic utilization | Used by 16-bit KOM multiplier | Used by 32-bit KOM multiplier | Used by 32-bit Baugh Wooly Multiplier | Used by 32-bit Dadda multiplier |
|---|---|---|---|---|---|
| 1 | No of slice registers | 5184 | 25596 | 6129 | 0 |
| 2 | No of slice LUT | 16632 | 53,271 | 70,443 | 55080 |
| 3 | No of fully used LUT FF pairs | 4320 | 25,596 | 1809 | 0 |
| 4 | No of bonded IOBs | 1755 | 3483 | 3699 | 3456 |

Table 2: Multiplication of 5x5 with another 5x5 matrix

| Sl no | Logic utilization | Used by 16-bit KOM multiplier | Used by 32-bit KOM multiplier | Used by 32-bit Baugh Wooly Multiplier | Used by 32-bit Dadda multiplier |
|---|---|---|---|---|---|
| 1 | No of slice registers | 24000 | 118,500 | 28375 | 0 |
| 2 | No of slice LUT | 77,000 | 246,625 | 326,125 | 255,000 |
| 3 | No of fully used LUT FF pairs | 20,000 | 118,500 | 8375 | 0 |
| 4 | No of bonded IOBs | 8125 | 16125 | 17,125 | 16000 |

Table 3: Multiplication of 7x7 with another 7x7 matrix

| Sl no | Logic utilization | Used by 16-bit KOM multiplier | Used by 32-bit KOM multiplier | Used by 32-bit Baugh Wooly Multiplier | Used by 32-bit Dadda multiplier |
|---|---|---|---|---|---|
| 1 | No of slice registers | 65856 | 325164 | 77861 | 0 |
| 2 | No of slice LUT | 211,288 | 676,739 | 894,887 | 699,720 |
| 3 | No of fully used LUT FF pairs | 54,880 | 325,164 | 22,981 | 0 |
| 4 | No of bonded IOBs | 22,295 | 44247 | 46991 | 43,904 |

Table 4: Multiplication of 11x11 with another 11x11 matrix

| Sl no | Logic utilization | Used by 16-bit KOM multiplier | Used by 32-bit KOM multiplier | Used by 32-bit Baugh Wooly Multiplier | Used by 32-bit Dadda multiplier |
|---|---|---|---|---|---|
| 1 | No of slice registers | 255,552 | 1261788 | 302137 | 0 |
| 2 | No of slice LUT | 819896 | 2626063 | 3472579 | 2715240 |
| 3 | No of fully used LUT FF pairs | 212960 | 1261788 | 89,177 | 0 |
| 4 | No of bonded IOBs | 86515 | 171699 | 182,347 | 170,368 |

VGG16, and VGG19. To compare resource utilization for this operation, Table 1-4 displays the multiplication of two square matrices of orders 3, 5, 7, and 11. The Karatsuba-Ofman multiplier has been identified as the most efficient method for this operation, utilizing the fewest slice LUTs. The study has shown that using this multiplier on FPGA can result in a powerful and cost-effective solution for accelerating the performance of deep neural networks in various signal processing applications, such as image and speech recognition, target detection, navigation, and cognitive radio.

Table 5: Comparison between digital Multiplier in terms of delay, Power dissipation

| SL NO | PARAMETER | KOM MULTIPLIER (32 BIT) (32-bit pipelined high speed, area optimized Karatsuba–Ofman multiplier) | KOM MULTIPLIER (16 BIT) (16-bit pipelined high speed, area optimized Karatsuba–Ofman multiplier) | Baugh Wooly Multiplier (32 bit) | Dadda multiplier (32 bit) |
|---|---|---|---|---|---|
| 1 | TIME DELAY | 4.604ns | 4.052ns | 15.415ns | 47.500ns |
| 2 | POWER DISSIPATION | 90.37 mW | 85.14 mW | | |

## VI. CONCLUSION

In conclusion, the investigation into FPGA-based Convolutional Neural Networks using the Karatsuba-Ofman digital multiplier revealed that it has significant potential for optimizing performance. This is due to its ability to use fewer slice LUTs and achieve lower delays when compared to other multipliers. Additionally analysis of popular Convolutional neural network architectures has shown that the Karatsuba-Ofman multiplier is the most effective for multiplying two square matrices of the same size, which is a fundamental operation in CNNs. These findings suggest that the Karatsuba-Ofman multiplier on FPGA can offer a cost-effective solution for improving the performance of deep neural networks in various signal processing applications, such as image and speech recognition, target detection, navigation, and cognitive radio. Overall, our study highlights the potential of using the Karatsuba-Ofman multiplier to address the memory bottleneck in FPGA-based CNN hardware accelerators and pave the way for more efficient and effective deep learning hardware architectures.

REFERENCES

[1] D. T. Kwadjo, J. M. Mbongue and C. Bobda, "Performance Exploration on Pre-implemented CNN Hardware Accelerator on FPGA," 2020 International Conference on Field-Programmable Technology (ICFPT), Maui, HI, USA, 2020, pp. 298-299, doi: 10.1109/ICFPT51103.2020.00055.

[2] H. P. Nghi and T. N. Thinh, "A CNN-Based Vehicle Identification Solution in Parking System With Hardware Accelerator on FPGA," 2022 RIVF International Conference on Computing and Communication Technologies (RIVF), Ho Chi Minh City, Vietnam, 2022, pp. 518-523, doi: 10.1109/RIVF55975.2022.10013847.

[3] W. Huang et al., "FPGA-Based High-Throughput CNN Hardware Accelerator With High Computing Resource Utilization Ratio," in IEEE Transactions on Neural Networks and Learning Systems, vol. 33, no. 8, pp. 4069-4083, Aug. 2022, doi: 10.1109/TNNLS.2021.3055814.

[4] A. J. A. El-Maksoud, M. Ebbed, A. H. Khalil and H. Mostafa, "Power Efficient Design of High-Performance Convolutional Neural Networks Hardware Accelerator on FPGA: A Case Study With GoogLeNet," in IEEE Access, vol. 9, pp. 151897-151911, 2021, doi: 10.1109/ACCESS.2021.3126838.

[5] Krizhevsky, Alex, Ilya Sutskever, and Geoffrey E. Hinton. "Imagenet classification with deep convolutional neural networks." Communications of the ACM 60, no. 6 (2017): 84-90.

[6] Simonyan, Karen, and Andrew Zisserman. "Very deep convolutional networks for large-scale image recognition." arXiv preprint arXiv:1409.1556 (2014).

[7] Manno, Steven V., "The Application of Systolic Architecture m VLSI Design, "University of Central Florida, 1986.

[8] Kung, H. T. "Why Systolic Architectures?" Computer, January 1982, pp. 37-46.

[9] J. G. Nash, "Computationally efficient systolic architecture for computing the discrete Fourier transform," in IEEE Transactions on Signal Processing, vol. 53, no. 12, pp. 4640-4651, Dec. 2005, doi: 10.1109/TSP.2005.859216

[10] Kung, H. T. (1979) Let's Design Algorithms for VLSI Systems. In: Proceedings of the Caltech Conference On Very Large Scale Integration. California Institute of Technology, Pasadena, CA, pp. 65-90.

[11] A. Karatsuba and Y. Ofman, "Multiplication of multidigit numbers on automata," Sov. Phys. Doklady, vol. 7, no. 7, pp. 595–596, 1963.

[12] K. Babulu, M. Kamaraju, P. Bujjibabu and K. Pradeep, "Design and implementation of H/W efficient Multiplier: Reversible logic gate approach," 2015 International Conference on Communications and Signal Processing (ICCSP), Melmaruvathur, India, 2015, pp. 1660-1664, doi: 10.1109/ICCSP.2015.7322800.

[13] K. Babulu, M. Kamaraju, P. Bujjibabu and K. Pradeep, "Design and implementation of H/W efficient Multiplier: Reversible logic gate approach," 2015 International Conference on Communications and Signal Processing (ICCSP), Melmaruvathur, India, 2015, pp. 1660-1664, doi: 10.1109/ICCSP.2015.7322800.

[14] L. Li and S. Li, "High-Performance Pipelined Architecture of Elliptic Curve Scalar Multiplication Over GF($2^m$)," in IEEE Transactions on Very Large Scale Integration (VLSI) Systems, vol. 24, no. 4, pp. 1223-1232, April 2016, doi: 10.1109/TVLSI.2015.2453360.